# Training neural nets to learn reactive potential energy surfaces using interactive quantum chemistry in virtual reality

Silvia Amabilino,[1,2,†] Lars A. Bratholm,[1,2,†] Simon J. Bennie,[1,2] Alain C. Vaucher,[3] Markus Reiher,[3] and David R. Glowacki[1,2,4*]

[†]these authors contributed equally this work
[1]School of Chemistry, University of Bristol, Bristol, BS8 1TS, UK; [2]Intangible Realities Laboratory, University of Bristol, BS8 1UB, UK; [3]ETH Zurich, Laboratory of Physical Chemistry, Zurich, Switzerland; [4]Department of Computer Science, University of Bristol, BS8 1UB, UK

*glowacki@bristol.ac.uk

**Abstract**

Whilst the primary bottleneck to a number of computational workflows was not so long ago limited by processing power, the rise of machine learning technologies has resulted in an interesting paradigm shift, which places increasing value on issues related to *data curation* – i.e., data size, quality, bias, format, and coverage. Increasingly, data-related issues are equally as important as the algorithmic methods used to process and learn from the data. Here we introduce an open source GPU-accelerated neural network (NN) framework for learning reactive potential energy surfaces (PESs), and investigate the use of real-time interactive *ab initio* molecular dynamics in virtual reality (iMD-VR) as a new strategy which enables human users to rapidly sample geometries along reaction pathways which can subsequently be used to train NNs to learn efficient reactive PESs. Focussing on hydrogen abstraction reactions of CN radical with isopentane, we compare the performance of NNs trained using iMD-VR data versus NNs trained using a more traditional method, namely molecular dynamics (MD) constrained to sample a predefined grid of points along the hydrogen abstraction reaction coordinate. Both the NN trained using iMD-VR data and the NN trained using the constrained MD data reproduce important qualitative features of the reactive PESs, such as a low and early barrier to abstraction. Quantitative analysis shows that NN learning is sensitive to the dataset used for training. Our results show that user-sampled structures obtained with the quantum chemical iMD-VR machinery enable excellent sampling in the vicinity of the minimum energy path (MEP). As a result, the NN trained on the iMD-VR data does very well predicting energies which are close to the (MEP), but less well predicting energies for 'off-path' structures. The NN trained on the constrained MD data does better predicting high-energy 'off-path' structures, given that it included a number of such structures in its training set.

# 1. Introduction

Over the past few years, statistical data inference in the form of machine learning has found application in a range of domains, driven in part by the fact that ubiquitous computational devices are producing an unprecedented amount of data. Alongside those developments, high-performance parallel computing architectures, along with easy-to-use software frameworks, are evolving to cope with analysis of these massive quantities of data. Machine learning has been used to carry out a wide range of tasks, including image recognition[1], strategy games[2], speech recognition[3], language translation[4], guiding consumer behaviour[5], etc. The molecular and material sciences have been no exception[6]. For example, machine learning is being touted as a potentially transformative technology in a range of domains, including proposing candidate drug molecules[7], developing organic semiconductors[8], planning synthetic chemistry strategies[9], and analysing molecular dynamics data[10].

One of the earliest applications for machine learning within the molecular sciences involves making accurate predictions of molecular energies and forces. This has been a longstanding interest for several workers in molecular science, particularly with the rise of molecular dynamics as a tool for furnishing microscopic insight, and its requirement for an enormous number of accurate evaluations of the energies and forces acting on all atoms. For relatively small systems, direct *ab initio* molecular dynamics can often be used to propagate dynamical equations of motion. However, for larger systems such methods become too computationally expensive. Over the years, a variety of methods have been proposed where non-linear functional forms are systematically combined so as to fit potential energy surfaces, including permutationally invariant fitting[11], Shepard interpolation[12], and Multi-State Empirical Valence Bond (MS-EVB) theory.[13] More recently, methods like Kernel Ridge Regression (KRR),[14] Reproducing Kernel Hilbert Space (RKHS),[15] and Artificial Neural Networks (NNs)[16] have gained attention for their ability to efficiently provide molecular energies and forces of the sort required for molecular dynamics.

The rise of machine learning technologies across so many domains has resulted in an interesting paradigm shift, which has seen increasing value placed on the data *per se* – i.e., the strategies for gathering data, the size of the dataset, the quality of the data, the bias of data, the format of the data, and the regularity of the data. All of these are at least as important (and in some cases, perhaps *more* important) as the algorithmic methods used to process the data. Whilst the primary bottleneck to a number of computational workflows was not so long ago limited by processing power, machine learning has created a scenario where the quantity, quality, and formatting of data often represent the principle bottleneck. In this brave new world, devising good methods for curating data in hyperdimensional spaces *is as important* as devising good methods for processing data.

For potential energy surface fitting, the strategies one uses to gather data are key. Grid searches, where one constructs a dense grid of points which systematically cover all degrees of freedom, are preferable where possible, but unfeasible for all but the smallest systems.[16] For more complex cases, a number of techniques have been proposed, including:

- MD trajectories, which sample the most probable regions of configuration space, but which struggle to sample the transitions between minima owing to the rare event problem.[16] In some cases, workers have adjusted the time interval between sampling as a function of atomic acceleration,[17] in an attempt to obtain a more uniform density of points.

- Constrained MD, where one preselects a grid of points along important degrees of freedom, followed by MD simulations in which the specified degrees of freedom are constrained to the grid values.[18]

- Enhanced sampling MD, using methods (e.g., replica exchange molecular dynamics, meta-dynamics, BXD, etc.) designed to sample regions of the potential energy surface which are not often sampled during standard MD. In general, such methods involve biasing the potential along some reaction coordinate or raising the temperature.[19]

- Adaptive sampling schemes (e.g., the GROW scheme[20] and related methods[21]), which usually begin by selecting a set of molecular configurations for the reaction of interest (e.g., points along a minimum energy path), fitting predefined functional forms to capture the energies of these points, running a small number of MD trajectories, identifying those points along the trajectories prone to the greatest error, and updating the fitted potential to accommodate these error-prone points. Iterative stages of running MD trajectories and fitting are then undertaken until some pre-specified convergence criteria is met.

In what follows, we revisit some of our previous work to develop accurate PES representations for the reaction of CN radicals with hydrocarbons, in order to aid experimental interpretation. In this article, our primary focus is on proposing, demonstrating, and evaluating new technologies for generating data on reactive potential energy surfaces, and then fitting that data using accurate machine learning models. Previously, our approach has been to develop parallel multi-state EVB models which accurately describe reactive dynamics in both the gas phase and in condensed phases[13b, 22]. In that work, we examined the reaction dynamics of CN + $C_6H_{12}$ (cyclohexane) using high level electronic structure theory to carry out grid-based scans of PES points along the minimum energy path (MEP) of hydrogen abstraction reactions. Using a Levenberg-Marquardt non-linear least squares algorithm, we then optimized the MS-EVB diabatic curve parameters in order to fit the reactive PES slices. In this article, we focus on a larger and more challenging system – i.e., CN + isopentane, using new computational technologies which we have recently developed, namely: (1) GPU-accelerated neural networks (NNs) for fitting potential energy surfaces, and (2) a VR-enabled framework which enables users to interact 'on-the-fly' with real-time quantum mechanical molecular dynamics simulations.[23] Using the VR framework, the user is able to quickly apply real-time biasing forces to specific atoms in the simulation, and generate large quantities of data focussed on those regions of the PES in which they are interested.

To date, the use of NNs to generate PESs for chemical reactions has focused mostly on small systems of 3-4 atoms[17, 24]. To the best of our knowledge, there has been relatively little work applying NNs to larger open-shell reactive systems of the sort examined herein. In this paper, we outline a new GPU-accelerated NN software framework which we have recently developed and use it to fit a NN to the PES of a cyano radical reacting with isopentane, shown in Figure *1*.

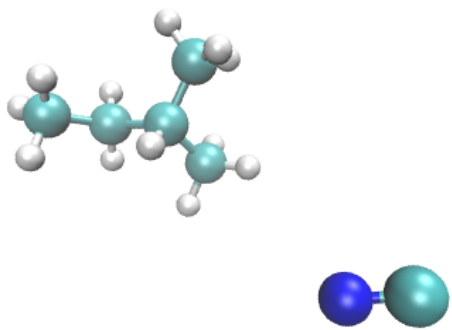

Figure 1: CN + isopentane

Compared to our previous work on CN + hexane, and CN + THF, this system offers an interesting test case on which to fit a NN PES. With 19 atoms, it represents (to the best of our knowledge) one of the largest radical systems where a NN-fitted potential has been developed. Moreover, the CN radical offers three classes of low-energy reaction pathways, where the CN abstracts hydrogens at either primary, secondary, or tertiary centers.

We have specifically been experimenting with NNs owing to the fact that: (1) Their functional form is not system specific; (2) advances in GPU acceleration allow NNs to be trained on large data sets; and (3) there are a number of readily available, well-supported community GPU-accelerated NN frameworks such as TensorFlow[25] and PyTorch.[26] In general, the quality of a NN-trained PES depends strongly on the quality of the data which is used for training. An interesting open question is the extent to which small sets of high-quality data vs. larger sets of sub-optimal data impact the accuracy of trained NN PES models. This is an especially important consideration in light of the fact that high quality electronic structure calculations like CCSD tend to be very expensive, and therefore it is important to explore effective strategies which might enable one to reduce the number of calls to expensive electronic structure theory methods. Our real-time VR-enabled quantum mechanical framework offers an easy way for bringing human intuition to bear so as to enable efficient sampling on hyperdimensional PESs, inspired by recent work examining how human agents undertake search tasks with hyperdimensional spaces[27]. In order to evaluate the VR-enabled search strategy, we undertake comparisons to NN-fitted PESs built from geometries sampled using constrained MD.

## 2. Computational Methods

### 2.1 Local Neural Networks

Broadly speaking, NNs that have been used to fit PESs fall into two categories – molecular NNs, and atomic NNs. Molecular NNs use a single feed forward NN to construct a direct functional relation between the molecular configuration and the potential energy[28], effectively encoding information about the entire system in a single vector. Molecular NNs can yield very accurate energies, but are less easily transferable to different systems, because adding new atoms to the system effectively requires 'rewiring the network',[28] as new input nodes need to be added and the model then needs to be trained. Atomic NNs solve both of these problems. Using atomic approaches, the potential energy ($E_{tot}$) is decomposed into $N$ atomic energy contributions ($E_i$), where $N$ is the number of atoms in the system:

$$E_{tot} = \sum_{i=1}^{N} E_i \qquad (1)$$

Using Eq 1, an atomic feed forward NN is effectively used to learn the relation between the atomic environment in the vicinity of a given atom, and its decomposed energy. To use Eq (1) in practice, each atom and its local chemical environment is converted to a vector. All vectors describing the environment around a specific element type will be input into the same feed forward NN, which will output its atomic energy. All the atomic energies are then summed together to obtain the total energy of the system. For example, Figure *2* shows a schematic atomic NN approach for representing the PES of the HCN molecule. Figure *2* shows how adding additional atoms to the system does not require changing the structure of the component NNs which are used to represent the molecular energy, and also illustrates the fact that the computational scaling with system size is linear so long as the long-range interactions are screened. The use of atomic NNs enables training carried out on smaller molecular systems to then be extended to making accurate energy predictions in considerably larger molecular systems, an area where we have obtained encouraging preliminary results. Since we intend to eventually simulate the reactions of CN with much larger hydrocarbons than isopentane, the local representations are more appropriate for our purposes. More in depth descriptions of these two types of NNs can be found in recent reviews.[28]

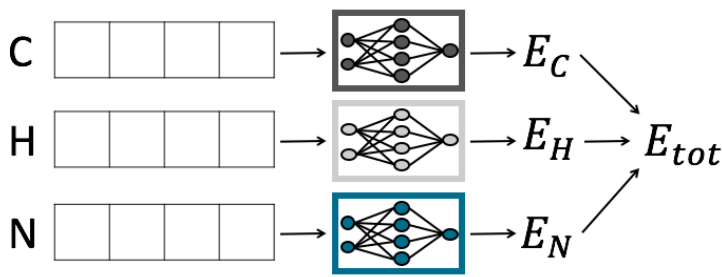

Figure 2: Representation of an atomic NN to calculate HCN. A feed forward NN is used to calculate the atomic energy for each atom embedded in the environment of every other atom. The atomic energies are then summed to give the total energy.

Over the last few years, software implementations for using atomic NNs to represent PESs have become available. For example, we are aware of Schnet[29], TensorMol[30], ANI[31], AMP[32] and RuNNer[33]. For the purposes of this work, we elected to develop our own implementation, as a result of the fact that the other packages either: (1) lacked GPU-acceleration, (2) were closed source, or (3) were not readily compatible with scikit-learn,[34] which is an extremely flexible and open community software package supporting a wide range of machine learning methods. For training the models described herein, we use Osprey[35] to carry out hyper-parameter optimization, which requires scikit-learn compatible models. We note here that we have made the NN implementation described in this article available through the open-source Quantum Machine Learning (QML) package[36].

*2.2 Atom-Centered Symmetry Function Representations*

Amongst the most important aspects of machine learning a PES involves choosing a suitable representation (i.e., input format) for how the machine learning algorithm processes the molecular structure. It is this representational difficulty which is principally responsible for the enormous proliferation in PES-fitting ML papers over the years. A good representation should be invariant to molecular translation and rotation, as well as permutation of atoms of the same element.[37] The conversion from Cartesian coordinates to the representation should be fast and the representation needs to be differentiable with respect to atomic positions so that forces can be calculated when needed.[16] The choice of representation very much depends on whether one is using a global molecular NN or a local atomic NN. Some commonly used global representations include (but are not limited to) the Coulomb matrix[38] and Spectrum of London and Axilrod-Teller-Muto (SLATM)[39]. Given that we have chosen to work with local atomic representations in this article, we have implemented Atom-Centered Symmetry Functions (ACSFs),[33] which includes multiple vectors, each of which describes the local environment around a given atom in the system. In this work, we use the modified Atom-Centered Symmetry Functions as formulated by Smith et al,[31] which are described by the following equations:

$$g_2 = \sum_{j \neq i} e^{-\eta(R_{ij}-R_s)^2} f_c(R_{ij}) \qquad (2)$$

$$g_3 = 2^{1-\zeta} \sum_{j \neq i, j \neq k} (1 + \cos(\theta_{ijk} - \theta_s))^\zeta \times e^{-\eta\left(\frac{R_{ij}+R_{ik}}{2}-R_s\right)^2} f_c(R_{ij}) f_c(R_{ik}) \qquad (3)$$

$$f_c(R_{ij}) = \begin{cases} 0.5 \left(\cos\left(\frac{\pi R_{ij}}{R_c}\right) + 1\right) & \text{for } R_{ij} \leq R_c \\ 0 & \text{for } R_{ij} > R_c \end{cases} \qquad (4)$$

where $g_2$ is the two-body symmetry functions, $g_3$ is the three-body symmetry functions and $f_c$ is a damping function. $\eta$ and $R_s$ determine respectively the width and location of the radial basis functions and $R_{ij}$ is the distance between atom $i$ and atom $j$. $\zeta$, $\theta_s$ determine the width of the angular basis, $\theta_{ijk}$ is the angle between atoms $i, j$ and $k$, $R_{ik}$ is the distance between atoms $i$ and $k$ and $R_c$ is the cut-off radius. We decided to use this formulation of the symmetry functions because they have been previously been shown to give good results[30-31], their functional form is clearly described, and they are better suited to probe the angular environment around each atom compared to the original formulation.[31] All the hyper-parameters ($\eta$, $R_s$, $R_c$, $\zeta$, $\theta_s$) need to be optimized to give the best possible result. $R_s$ and $\theta_s$ indicate where the basis functions are centered; typically an equidistant grid of distance-angle pairs is used to create an array of symmetry functions that fully captures the local environment. We reduced the number of hyper-parameters by requiring that $N_{basis}$ radial basis functions were placed equidistant between 0.8 Å and the cut-off radius $R_C$, and that $N_{basis}$ angular basis functions were placed equidistant between 0 and $\pi$. To both de-correlate and further reduce the number of hyper-parameters, we re-parametrised $\eta$ and $\zeta$ to be a function of a precision parameter $\tau$ that specifies the overlap between the basis functions. We define $1/\tau$ to be the function value of the intersection of two neighbouring radial functions and $2/\tau$ to be the function value of the intersection of two neighbouring angular functions. Then, $\eta$ and $\zeta$ can be expressed as functions of $\tau$:

$$\eta = \frac{4 \log(\tau)(N_{basis}-1)^2}{(R_c - r_{min})^2} \qquad (5)$$

$$\zeta = -\frac{\log(\tau)}{2 \cdot \log\left(\cos\left(\frac{\pi}{4N_{basis}-4}\right)\right)} \qquad (6)$$

where $r_{min}$ is the distance at which to start placing radial basis functions, that we have set to 0.8 Å in this study. We refer to the SI for further details.

The atomic NNs and ACSFs[31, 33] outlined above have been implemented in the QML (Quantum Machine Learning) Python package, a project initiated by von Lilienfeld and co-workers. The aim of QML is to provide user-friendly and efficient implementations of molecular representations and machine learning models for describing the properties of molecules and solids. The NNs are implemented fully in TensorFlow[25], which makes GPU acceleration of the NN training straightforward. We have implemented the ACSFs in both Fortran and TensorFlow. The TensorFlow implementation benefits from vectorisation and GPU acceleration, while the Fortran implementation uses OpenMP[40] to parallelise loops. The advantage of the TensorFlow implementation is that one can obtain the gradients of the energy with respect to the Cartesian coordinates with no extra effort. It also makes it possible to include forces in the training of the NN. Because we did not require forces for the purposes of this article, we utilized the Fortran implementation to obtain the results described herein. The Fortran implementation generates about 800 ACSF representations per second, while the TensorFlow implementation generates about 400 ACSF representations per second. We also contributed a scikit-learn[34] interface to QML, enabling its various machine learning models to be used in the same way as any other model available through scikit-learn. This was done to make the code compatible with packages such as Osprey[35] for hyper-parameter optimisation. Instructions and accompanying scripts for how to use the NN/ACSF implementation within QML are available in the supporting information.

## *2.3 Hyper-parameter optimization*

The shape of the neural network, regularization strength and parameters of the Atom-Centered Symmetry Functions were optimized with Gaussian processes (GP) within the Osprey software package. The Expected Improvement (EI)[41] and Upper Confidence Bound (UCB)[42] acquisition functions were used in parallel to explore the search space. This was done separately for the structures sampled with VR and constrained MD. The procedure which we used to optimize the network hyperparameters is as follows:

1. We fitted a GP to the MAE as a function of hyperparameters. The GP fit to the MAE has a variance associated with it which indicates the uncertainty of the GP prediction. We chose between a Gaussian, Matern 3/2, and Matern 5/2 kernel, depending on which gave the highest likelihood. We used Akaike's information criterion with a correction for small sample size (AICc)[43] to reduce the number of available kernels.
2. We similarly fit a GP to the mean time required to train the network using the same procedure discussed above.
3. To optimize the hyper-parameters, we chose the values that minimized the MAE plus one standard deviation on the fitted GP hyperparameter surface. The idea here was to select good hyperparameter values with a high degree of confidence – i.e., a low mean and variance of the GP-predicted MAE. In addition, we avoided hyperparameters which required the model to train for a very long time. We did this by adding a constraint where we did not allow hyperparameters giving rise to models which the GP predicted would take more than 8 hours to train.

Instructions on how to replicate the hyper-parameter optimization in Osprey are available in the supporting information.

*2.4 Interactive quantum chemistry in VR for sampling reaction pathways*

Recent work by O'Connor *et al*[23] describes a multi-user VR-enabled interactive molecular dynamics framework, which combines rigorous real-time atomistic physics simulations with commodity VR hardware, which we have recently made available as an open-source software package hosted at [www.gitlab.com/intangiblerealities](www.gitlab.com/intangiblerealities). The framework, which we call 'Narupa', allows users to visualize and sample, with atomic-level precision, the structures and dynamics of complex molecular structures "on the fly" and to interact with other users in the same virtual environment. In a series of controlled studies, we quantitatively demonstrated that users within an interactive VR environment could complete sophisticated molecular modeling tasks significantly faster than they could using conventional interfaces, especially for molecular pathways and structural transitions whose conformational choreographies were intrinsically three-dimensional.

The Narupa framework relies on a client/server model, in which an HTC Vive VR client is connected to a server (hosted either on a local compute cluster or on a cloud supercomputer). The server hosts a user-specified force engine, which runs a real-time molecular dynamics simulation, and which streams the results in real-time to connected clients. Virtual reality clients render the MD simulation results, which are continually updated for users to see. As the simulation is running, users are able to literally reach into the simulation and manipulate both individual atoms and groups of atoms, so as to bias their dynamics 'on-the-fly'. The ability to achieve such biasing is implemented as an external force field (whose strength the user may control) which is integrated into the MD simulation. This framework enables rapid and intuitive sampling of configurations for a given molecular system in order to rapidly test hypotheses and generate dynamical pathways.

The flexibility of the Narupa framework arises from its modularity. For example, it exposes a plug-in interface which enables new force engines to be deployed without modifying the original source code, encouraging extensibility. In our previous work, we have primarily focussed on applications which used real-time molecular mechanics force engines. For the purposes of this article, where we are concerned with sampling bond-breaking and making, we have utilized the Narupa API to enable communication with two quantum mechanical force engines: (1) an API-compatible wrapper enabling communication between Narupa and the tight-binding DFTB+[44] package, which required us to refactor the program so that it could be called as a library; and (2) the SCINE Sparrow package developed by Reiher and co-workers[45] ([http://scine.ethz.ch](http://scine.ethz.ch)), which includes implementations of tight-binding engines like DFTB alongside a suite of other semi-empirical methods.[46] To obtain the results described herein, we have utilized the SCINE Sparrow implementation of PM6, which we found to give better energy pathways than DFTB (with the mio parameter set) for the particular system under investigation herein. Using real-time PM6 in VR, we were able to sample a wide range of H-abstraction pathways at the primary, secondary, and tertiary sites on isopentane. Figure 3 and supplementary video 1 (available also at www.vimeo.com/311438872) show a representative user-guided abstraction pathway, obtained by bringing the CN into proximity with a primary hydrogen on isopentane. To the best of our knowledge, this article represents the first report of real-time quantum mechanical force engines being steered using an interactive VR environment.

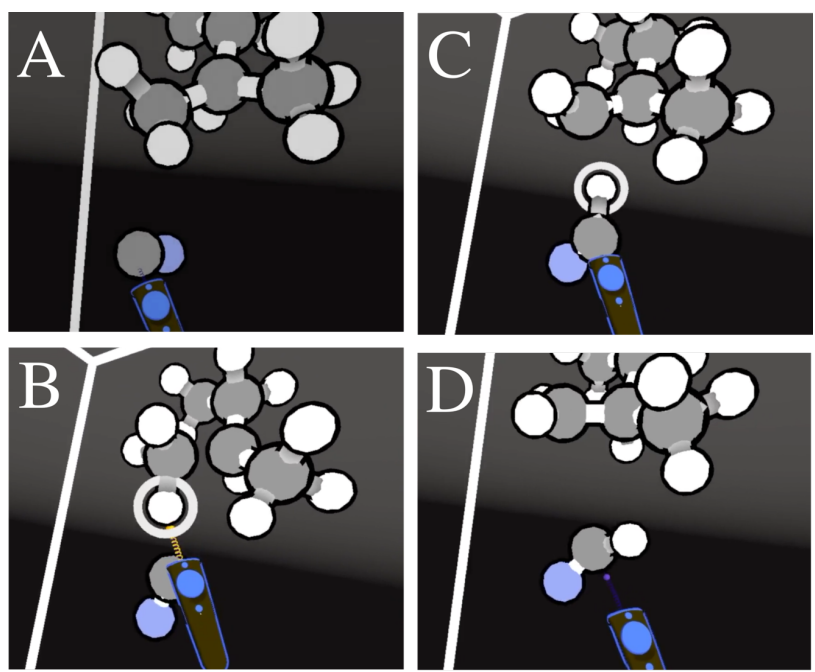

Figure 3: first person view of a user within VR as they utilize the Narupa software to interactively sample reaction pathways. The blue objects that can be seen are the wireless controllers, which enable a user to reach into the system and carry out atom-specific manipulation of molecular systems. The white highlight on a particular atom (e.g., the hydrogen undergoing transfer) indicates the fact that an external force is being applied by the user on that particular atom. Panel (A) shows the CN + isopentane reactants; panels (B) and (D) show the user exerting a force on a tertiary hydrogen to enact transfer, and panel (D) shows the HCN + isopentyl radical products

To undertake accelerated sampling of CN + isopentane reaction pathways in VR, a starting structure of each of the reactants in XYZ format was loaded into the NarupaXR environment, spawned in random (non-overlapping) positions within a cubic box with length 30 Å. Real-time MD simulations were run using a Velocity Verlet integrator with a time step of 0.5 fs. An Andersen thermostat was used to maintain the system temperature at 300 K, with a collision frequency of 10 ps$^{-1}$. The system was constrained to stay within the box via velocity inversion, ensuring the reactants were within reach of the user without the need to use periodic boundaries. For interaction, we utilized a spring potential with a force constant of 1000 kJ/(mol*a.m.u). We used a velocity re-initialisation procedure to rapidly re-equilibrate the system between interactions, removing momentum which users introduced into atoms during the course of an interaction (e.g., the atom undergoing transfer in Figure *3*). The SCINE implementation of PM6 was used, with the default set of parameters. For visualisation purposes, a dynamic bonds algorithm was used which, at every step, generates the current set of bonds using a simple distance criterion. In this study a length of 1.4 Å was used to define a bond. In order to sample hydrogen abstractions, the reactants were brought in proximity to enable the reaction to take place, and the nascent products were then moved away from each other, as shown in supplementary video 1. Throughout this process, conformations of the system were logged to an XYZ file every step.

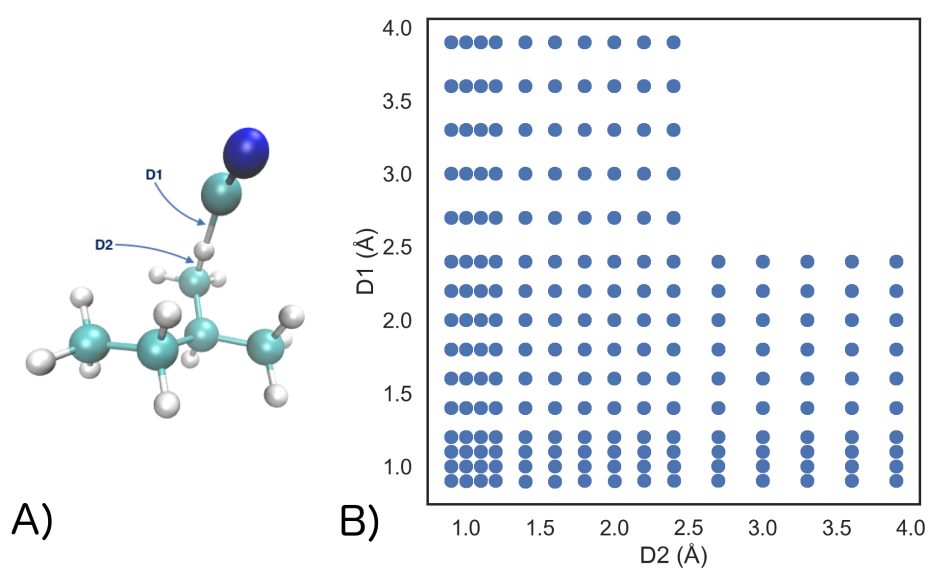

Figure 4: panel (A) shows the D1 and D2 degrees of freedom used to constrain MD simulations in order to sample a primary abstraction mechanism in CN + isopentane. The grid in panel (B) shows the constrained values of D1 and D2 used to carry out constrained MD sampling.

## *2.5 Constrained Optimizations & Constrained MD*

To generate the second data set, we used Constrained Molecular Dynamics (CMD). We chose the CH distance on the Isopentane and on the Cyano radical as the degrees of freedom to constrain (Figure *4*) and constrained each of the 12 hydrogens in turn. We performed these constrained MD simulations using PM6[47] in the CP2K package[48]. The simulations were run in the NVT ensemble at 300 K with the CSVR thermostat[49], using a 20 Å simulation box. We used a time step of 1 fs for a total of 5,000 steps, with a structure being logged to an XYZ file every 500 steps. The values of the constraints which we used are shown in Figure *4*B.

We also generated a minimized PES to be used as a test for the trained neural networks. To construct the minimized PES, we selected structures with the same constraints mentioned above, but all structures were optimised with CF-uPBE0/SVP. Again, the distance constraints depicted in **Error! Reference source not found.** were used for a single primary hydrogen, but constraints larger than 3 Å were neglected, as they proved difficult to converge. Additionally, some optimizations resulted in a non-constrained hydrogen being abstracted. These optimizations were stopped early and removed. Input files for both optimizations and constrained MD are available in the SI.

*2.6 Higher level electronic structure calculations*

The VR-enabled interactive PM6 calculations enabled us to quickly sample a range of important structures along the CN + Isopentane hydrogen abstraction pathway. To refine the PM6 energies from the VR and constrained MD simulations, we carried out subsequent electronic structure calculations using MOLPRO[50] to compute coulomb fitted[51] unrestricted PBE[52] with the Def2-TZVP[53] basis set (henceforth referred to as CF-uPBE/TZVP). This method was chosen because the reaction enthalpy of a primary and secondary hydrogen abstraction matched well with experimental values[54]. The computed reaction enthalpy on CF-uPBE0/SVP[55] optimized structures were -101.9 kJ mol$^{-1}$ and -117.4 kJ mol$^{-1}$ for primary and secondary hydrogen abstractions respectively, compared to the experimental values of -108.8 kJ mol$^{-1}$ and -121.3 kJ mol$^{-1}$.

## 3. Results & Discussion

As was mentioned in the introduction, the generation of the data set is key for any type of machine learning. In this work, we generated data using two different procedures. In what follows, we refer to the data generated using the interactive Narupa framework as the iMD-VR dataset. Data generated using constrained MD we refer to as the constrained MD dataset. In what follows, we will describe the sampling we achieved using both the iMD-VR dataset, and the constrained MD dataset, as well as the results which we obtained by training neural networks on each dataset in turn.

*3.1 The iMD-VR dataset*

Figure 5A shows the pruned data corresponding to the energetic profile of 19 CN-isopentane reactive trajectories which we generated using the Narupa VR framework. Initially about 25 trajectories were generated, which took approximately 1 h. Then, all those trajectories where the reaction pathway did not correspond to what we specifically set out to model for the purposes of this paper (e.g., formation of HNC, accidentally breaking the isopentane, etc.) were removed. This left us with 19 trajectories, 11 of which were primary abstractions, 3 of which were secondary and 5 of which were tertiary. These raw trajectories were then further pruned by keeping 600 configurations before and after each trajectory reaches a reference energy of 290.175 Ha, which corresponds to an energy which is approximately half way between that of the stable reactants and products. This left us with a total of 22,756 data points. The energies of the pruned trajectories were recalculated with CF-uPBE/TZVP. We then subtracted the arbitrary reference energy and removed all those structures with an energy of 150 kJ mol$^{-1}$ above this reference. This is well above the energy of the free reactants as shown in

Figure *5*B.

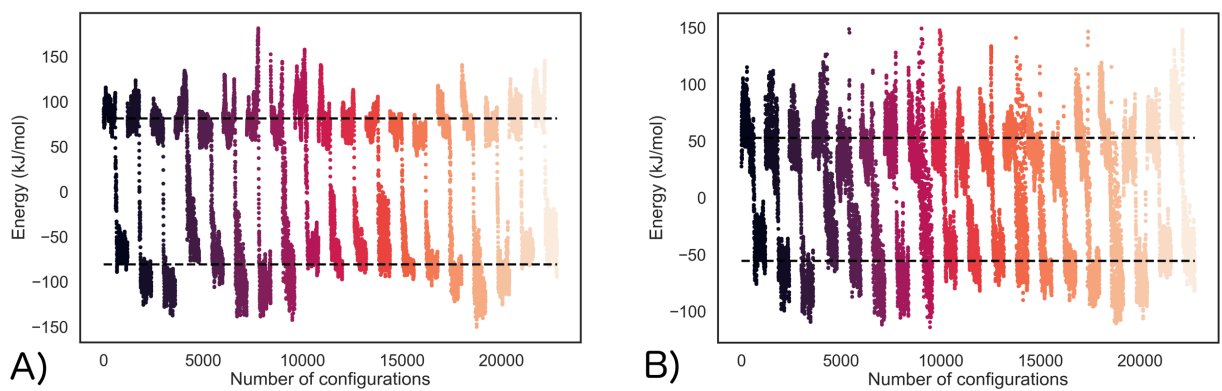

Figure 5: Trajectories obtained in VR of the CN radical reacting with isopentane. The configurations from each trajectory are coloured differently and the energy is from A) PM6 and B) CF-uPBE/TZVP level of theory. The dotted lines indicate the average energy of the first and the last 400 frames of each trajectory, corresponding to the average energy of the reactants and products, respectively.

The orange curve in Figure 6 shows the kernel density estimate (KDE) for how frequently geometries were sampled at specific energies using the iMD-VR approach. The KDE is constructed by placing a Gaussian on each observation and then summing all the Gaussians to obtain smooth histograms of how often each value occurs. Figure 6 and Figure 7 indicate that the iMD-VR approach samples the equilibrium structures more often than the transition regions as the bimodal density clearly corresponds to reactant and product structures. The peak corresponding to the product structures is slightly wider than the reactant peak. This is because the energy of the products resulting from primary, secondary and tertiary abstractions are all slightly different, whereas the reactant energies are largely the same. Compared to constrained MD sampling, Figs 6 and 7 clearly show that the iMD-VR approach enables better sampling in the vicinity of the minimum energy path (MEP), as a result of instabilities that arose during constrained MD sampling, which are discussed in further detail below.

### *3.2 The constrained MD dataset*

To obtain this data set, we chose two reaction coordinates as illustrated in Figure *4* for each of the 12 reactive hydrogens: (1) The distance between the CN carbon and each isopentane hydrogen; and (2) The distance between each isopentane hydrogen and the carbon to which it is bonded. These distances were varied as shown in Figure *4*. The spacing used between all the points is not equal. Closer spacing was used near the equilibrium distances of the reactant and product (CH distances between 0.9 Å and 1.2 Å) in order to get higher resolution along the MEP. Larger spacing was used for larger CH distances. For each point on the grid, a constrained MD simulation with the two CH distances constrained was carried out using PM6.

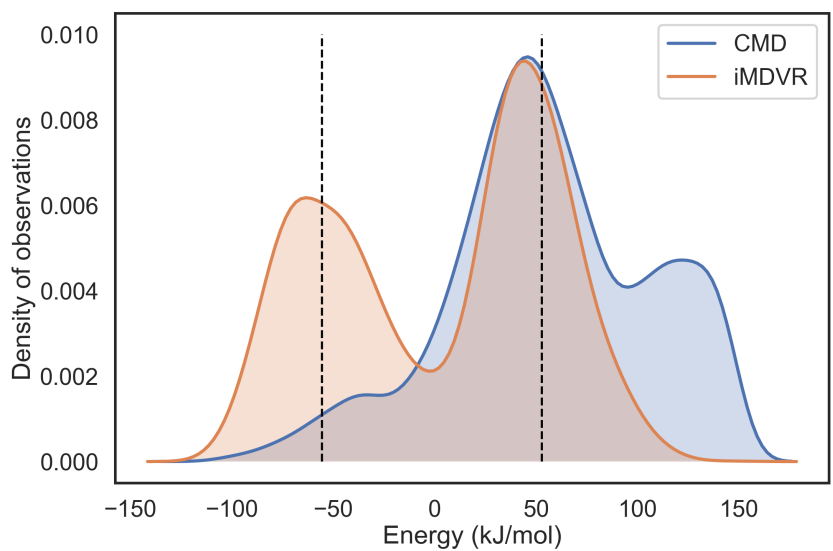

Figure 6: Kernel density estimate of the DFT energies of the configurations sampled using the iMD-VR approach (orange) and the constrained MD approach (blue). The dotted lines show the average energy of the reactants and the products as shown in Figure 5B.

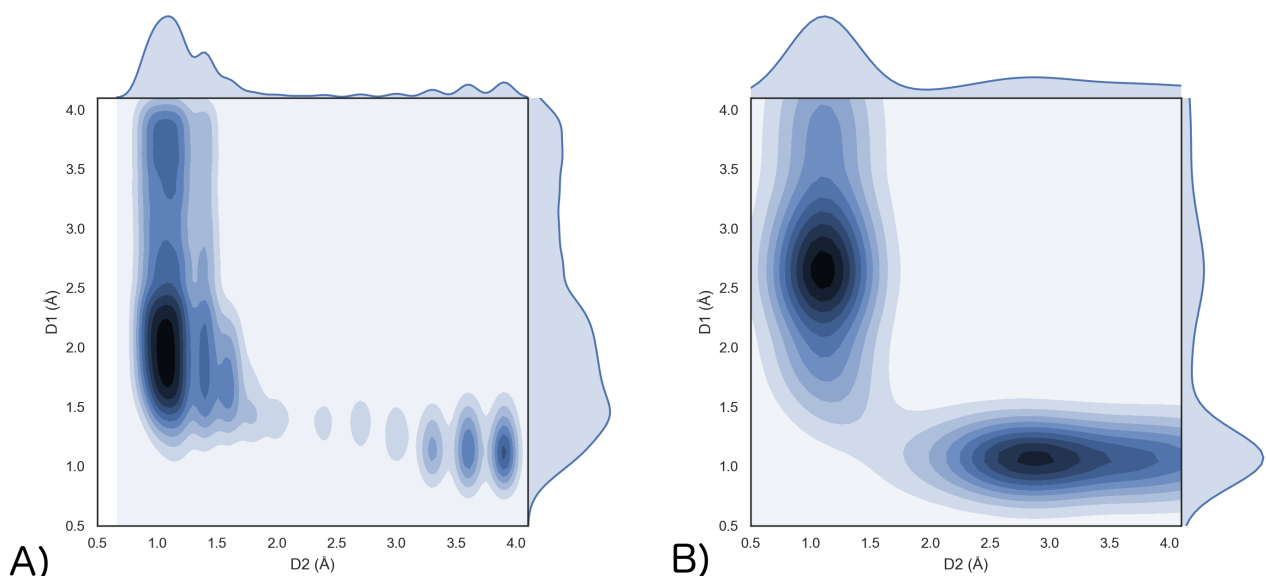

Figure 7: Kernel Density Estimation of the configurations sampled using constrained MD (A) and iMD-VR (B).

In many cases, enforcing the distance constraints led to a scenario where the molecules broke into multiple fragments. For example, several simulations resulted in a non-constrained hydrogen being abstracted or formation of $H_2$. We removed these spurious structures, which occur along a reaction pathway which did not correspond to what we specifically set out to model for the purposes of this paper, using similar logic as we did in pruning the data generated using iMD-VR. As illustrated in Figure *7*, structures in the product region of the configuration space were difficult to sample successfully using CMD. All structures with energies more than 150 kJ mol$^{-1}$ above the same reference energy used for the VR sampling (290.175 Ha) were also removed. After pruning the unwanted structures we were left with 7,621 structures out of the possible 24,000 (10 snapshots for 12 hydrogens and 200 constraints). These structures were then refined at the CF-uPBE/TZVP level. As can be seen from the blue kernel density estimate in Figure 6, the distribution of the constrained MD energies is different from the one obtained through VR sampling. The iMD-VR data set includes more configurations with energies around -50 kJ mol$^{-1}$ while the constrained MD data set contains a more significant fraction of higher-energy configurations, with values between 100 kJ mol$^{-1}$ and 150 kJ mol$^{-1}$. In the iMD-VR data set, the user guides the reaction along the minimum energy path (MEP). Figure 6 shows that the iMD-VR sampled structures do not stray far from the MEP, which is reflected in the population shown in Figure 7. This is different to the constrained MD data set, which includes sampling of higher energy structures in the vicinity of the transition state. Compared to the iMD-VR data set, the CMD data shows a noticeable lack of structures in the region of 2.0 Å < D2 < 3.0 Å, owing to instabilities in the electronic structure theory dynamics.

*3.4 Learning curves on iMD-VR dataset*

Before training a NN for predicting energies of our data sets, we made a learning curve with each data set. One trajectory of the abstraction of a primary H was removed from the data set, in order to use it as a separate test set. This left us with 21563 data points. From this data, we randomly selected 100, 300, 1,000, 3,000, 10,000 data points. For each sub-set of data and the 21,563 data points, we did 3-fold cross validation, meaning that a NN is trained on 2/3 of the data and tested on the remaining 1/3 and the process is repeated 3 times. This means that the NN was trained on 67, 200, 667, 2,000, 6,667 and 14,375 data points, respectively. The hyper-parameters were optimised only once on the data set with 14,375 data points due to the computational expense of the process. The learning curve is shown in orange in Figure *8*, where the mean absolute error (MAE) was evaluated on the trajectory not used for training, and the uncertainty is the standard deviation of the MAE obtained by each NN trained on the different folds of the data.

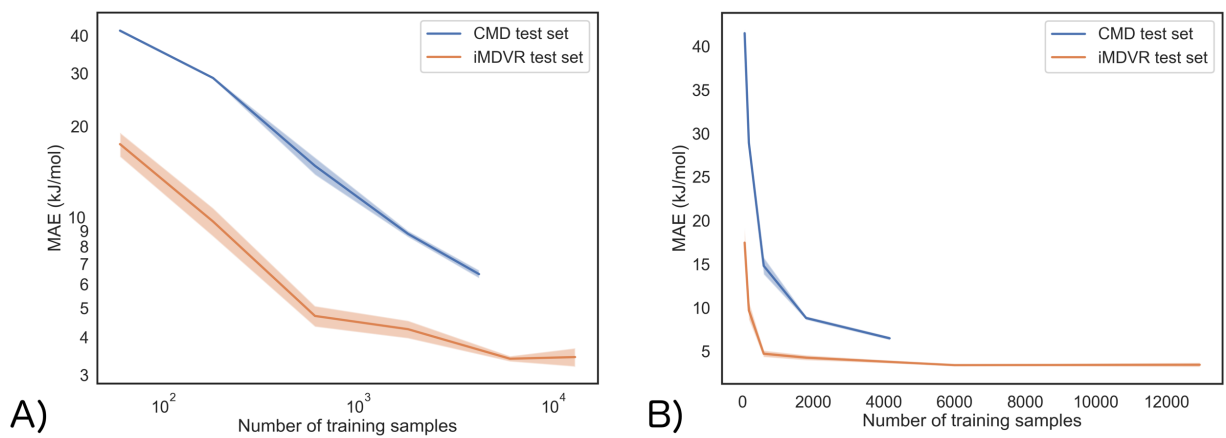

Figure 8: Log-log plot (A) and linear scale plot (B) of the learning curve for the atomic NN trained on the iMD-VR dataset (orange) and on constrained MD dataset (blue).

The learning curves in Figure 8 show that our ACSF NN implementation can accurately represent the energetics of hydrogen abstraction reactions on the CN-Isopentane PES. With increasing data, the error decreases, converging to $3.39 \pm 0.1$ kJ mol$^{-1}$ when training is conducted using 6,667 data points in the iMD-VR dataset. Figure *8* shows that doubling the number of data points in the iMD-VR data set used for NN training does not improve the error further.

### *3.4 Learning curves on the constrained MD dataset*

The same procedure was used for the constrained MD data set, except that this data set only contains 7,621 data points, so the learning curve could not be run as far. Here, a set of configurations where one primary hydrogen abstraction pathway is constrained was kept separate from the rest of the data, in order to use it as the test set. This left us with a total of 6,939 data points. The hyper parameters were again only optimised on the full data set. This learning curve is shown in blue in Figure *8*. Compared to the iMD-VR data set, the learning curve for the constrained MD dataset shows behaviour which is considerably more linear. When trained on 4,626 data points, the MAE for the constrained MD dataset is $6.5 \pm 0.2$ kJ mol$^{-1}$.

### *3.5 Training and cross predictions*

For both the iMD-VR and the constrained MD data set, two distinct NNs were trained on 7,621 data points. The model trained on iMD-VR data (referred to as iMD-VR-NN) was used to predict the energy of an abstraction trajectory sampled in VR, as well as the energies of 7,621 structures from the constrained MD data set (cross prediction). The opposite was done for the NN trained on the constrained MD data (referred to as CMD-NN). The MAE of the predictions and the cross-predictions are shown in Table *1*.

Table 1: Mean Absolute Error (kJ mol⁻¹) for test set predictions and cross predictions, where the error given in one standard deviation.

|  |  | Predicting on | |
|---|---|---|---|
|  |  | iMD-VR data | CMD data |
| Training on | iMD-VR data | 3.6 ± 5.0 | 12.4 ± 21.0 |
|  | CMD data | 6.4 ± 11.0 | 5.1 ± 6.0 |

Table *1* shows that the CMD-NN can predict with about the same accuracy the structures from both its test set and from the iMD-VR data set, which indicates that the external force applied by the user did not considerably distort the molecules into high-energy structures. When the CMD-NN predicts the low energy structures in the iMD-VR data, the MAE is larger compared to when the iMD-VR-NN predicts the lower energy structures in the constrained MD data set (Figure 9). This arises from the fact that user-sampled structures using the quantum chemical iMD-VR machinery appear to offer a more robust option for sampling points along the MEP (i.e., Figure 7A shows that constrained MD struggled to provide good MEP sampling in the region of 2.0 Å < D2 < 3.0 Å). The iMD-VR-NN does better than the constrained MD on its own test set by 1.5 kJ mol⁻¹, but it has an overall MAE that is over twice the size when predicting the constrained MD structures. Figure *9* shows the correlation plots for the cross predictions. For the iMD-VR-NN, the energies were divided in three regions. A MAE (kJ mol⁻¹) was calculated for each region. This shows that the iMD-VR-NN does better than the CMD-NN at predicting lower energy structures which are not far off the minimum energy path (MEP). However, since few of the iMD-VR sampled structures sampled the region between 100 and 150 kJ mol⁻¹, the errors in this energy range are very large. The errors are large also for structures with energy greater than 50 kJ mol⁻¹, where the iMD-VR data set had a considerable number of samples. This suggests that the constrained MD may have sampled different regions of the PES with similar energy, which is also observed in Figure 6 and Figure 7. Specifically, in for 'off-path' PES regions in the vicinity of the TS, the constrained MD data have regions of the PES which better sampled than the CMD approach – i.e., for the region where 1.5 Å < D2 < 2.0 Å and 1.5 Å < D1 < 2.0 Å, and also the region where 0.5 Å < D2 < 1.5 Å and 0.5 Å < D1 < 1.5 Å.

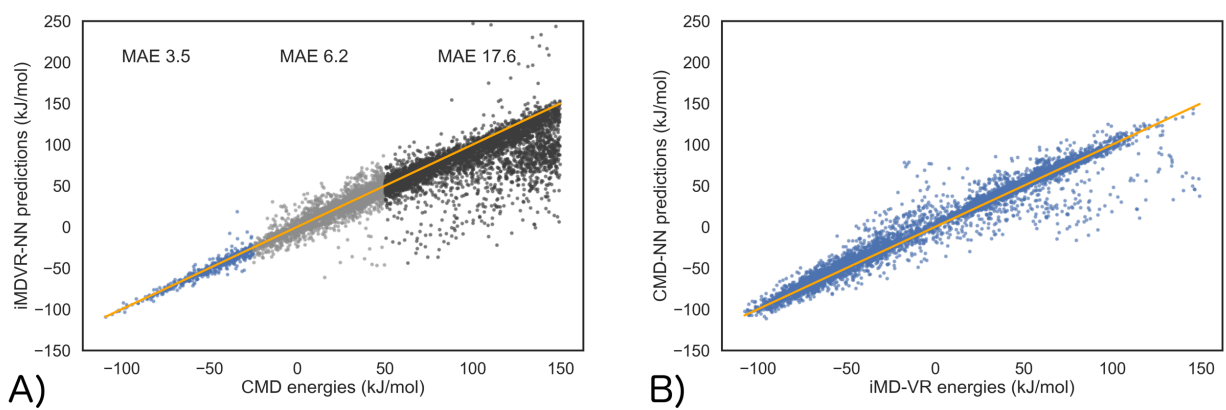

Figure 9: Comparison of models trained on the iMD-VR data set (A) and on the constrained MD data set (B). This shows the correlation plot for the cross-predictions, i.e. when the iMD-VR-NN predict the constrained MD energies (A) and when the CMD-NN predict the iMD-VR energies (B). For the iMD-VR-NN predicting the energies of the structures in the constrained MD data set (A), the energies were divided into three regions and a MAE (kJ mol$^{-1}$) was calculated for the predictions in each region.

## 4. PES Prediction

Relaxed potential energy surface scans were carried out at the CF-uPBE0/SVP level of theory with the same grid used for the constrained MD, where the two reaction coordinates considered were the distance of the hydrogen being abstracted to the isopentane carbon (D1) and to the cyano carbon (D2). All other degrees of freedom apart from D1 and D2 were optimised. The CF-uPBE/TZVP energies were interpolated to obtain a potential energy surface for the abstraction of a primary hydrogen as shown in Figure *10*. The surface was plotted only for structures with energies up to 100 kJ mol$^{-1}$ above the reference (290.175 Ha). As these structures are relaxed under constraints, the energies along the minimum energy path are much lower than those in the iMD-VR or constrained MD set, as shown in Figure 6 and Figure 7.

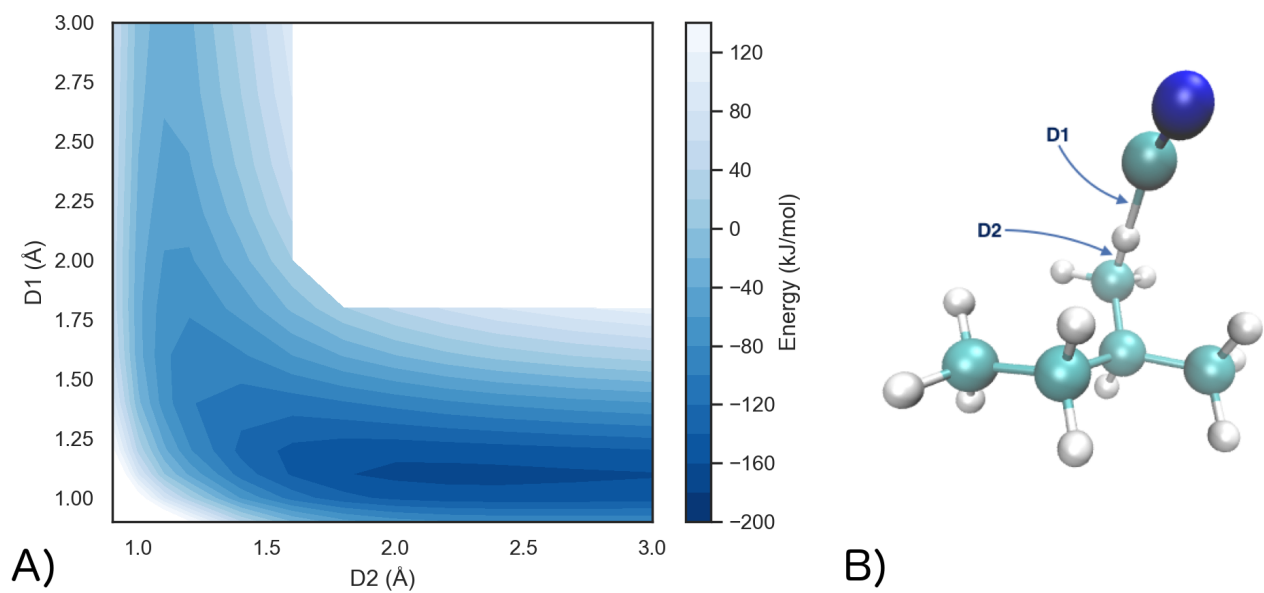

Figure 10: (A) Surface obtained from interpolating the relaxed potential energy scans for the abstraction of a primary hydrogen. (B) Explanation of the degrees of freedom constrained in order to make the plot in panel (A).

The iMD-VR-NN and the CMD-NN were used to predict the energies of the optimized structures. The results for the iMD-VR-NN and the CMD-NN predicting the surface are shown in Figure *11*. The difference between the predicted surfaces and the DFT surface are shown in Figure *12*.

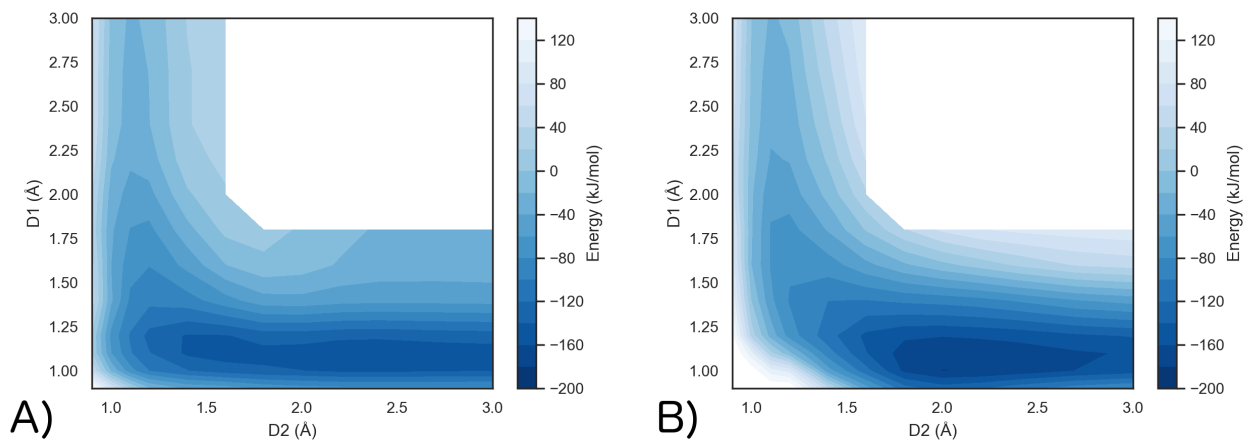

Figure 11: PES obtained by predicting the energies of the optimised structures using (A) the iMD-VR-NN and (B) the CMD-NN.

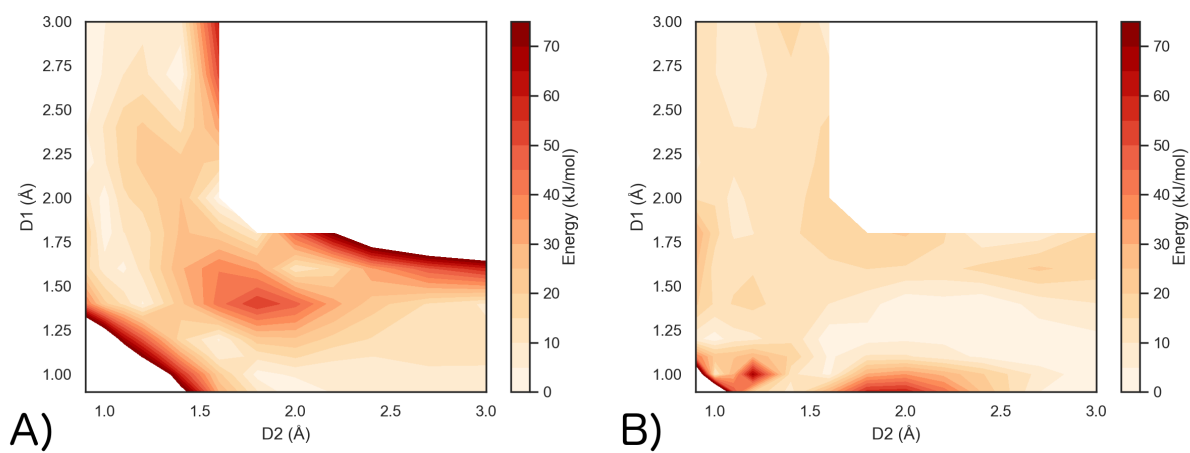

Figure 12: Difference between the iMD-VR-NN surface and the DFT surface (A) and difference between the CMD-NN surface and the DFT surface (B). Values where the error is larger than 80 kJ mol$^{-1}$ are not shown.

Figure *11* shows that both the iMD-VR-NN and the CMD-NN reproduce the qualitative features of the primary hydrogen abstraction surface reasonably well. Both indicate a reaction with an early transition state, and a negligible barrier to hydrogen abstraction.[22a] The CMD-NN reproduces the shape of the potential energy surface better than the iMD-VR-NN. The reason appears to derive from the fact that the constrained MD dataset was designed to sample more uniformly regions of higher energy (e.g., at small values of D1 and D2, as shown in Figure *4* and Figure *7*) than the iMD-VR dataset. The errors in the iMD-VR-NN are most significant in regions which are off the minimum energy reaction path (MEP), owing to the fact that users carrying out interactive molecular dynamics in VR tend to cluster their sampling in the region of the MEP, leaving 'off-path' regions relatively under-sampled by comparison. It remains to be seen the extent to which such regions are visited

during the course of MD trajectories run under typical experimental conditions, and this is an area which we plan to investigate in further work.

## 5. Conclusions

In this work, we introduced an open source GPU-accelerated neural net (NN) framework for learning reactive PESs using atom-centered symmetry function ACSF representations, applied it to study primary, secondary, and tertiary hydrogen abstraction reactions of CN radical with isopentane. To obtain the data required to train the NN, we investigated two different data-gathering approaches: (1) real-time *ab initio* iMD-VR as a new strategy enabling users to rapidly sample geometries along reaction pathways which can subsequently be used to train NNs to learn efficient reactive PESs, and (2) CMD, which was set up to sample a predefined grid of points along the hydrogen abstraction reaction coordinate. We used these approaches to generate two different data sets on which we trained two different NNs, and then compared the performance of the respective NNs trained on each data set. Qualitatively, both the NN trained using iMD-VR data and the NN trained using the constrained MD data reproduce important qualitative features of the reactive PESs, such as a low and early barrier to abstraction. Quantitative analysis shows that NN learning is sensitive to the dataset used for training. For example, the NN trained on the iMD-VR data does very well predicting energies which are close to the minimum energy path (MEP), but less well in predicting the energy of 'off-path' structures. This is because the structures sampled using the quantum chemical iMD-VR machinery do a much better job sampling the MEP than constrained MD. On the other hand, the NN trained on the constrained MD data does better in predicting the energy of higher-energy 'off-path' structures, given that it included a number of such structures in its training set.

These results show that user-sampled structures within the iMD-VR framework tend to be located not far off the minimum energy path, and that they sample the equilibrium structures more than the transition state regions. The constrained MD on the other hand enables better sampling of transition state regions and the higher energy structures, but samples less well low energy structures. This work suggests that it may be possible to construct a more optimal data-sampling strategy by combining user-sampled structures with automated approaches like CMD. For example, an algorithm which enables excursions from user-sampled structures along the MEP may allow us to efficiently access high-energy structures in important regions of configuration space.

This work clearly shows that user-biased forces within the quantum mechanical iMD-VR framework enable us to gather physically meaningful structures along an MEP. This is an important result, because it provides evidence that the iMD-VR framework might enable us to efficiently 'crowd-source' the gathering of data sets for important classes of chemical transformations, whose reaction pathways could then be fit using machine-learning technology. Moving forward, it will be interesting to explore the extent to which excursions from the MEP can be facilitated within the iMD-VR framework. For example, by changing the parameters of the thermostat to less quickly damp the interactive MD, it may be possible to obtain structures which make larger deviations from the MEP. In this work, we used only energies and geometries in our training set; however, we also hope to examine whether including forces alongside energies within the training data improves the ability of the NN to learn accurate PESs. In future work, we intend to use the machinery outlined herein to build reactive PESs for the reaction of CN with larger molecules, so that we can understand CN scattering experiments at liquid hydrocarbon surfaces.[56] By screening long range interactions, it should be possible to using the ACSFs to build accurate reactive PESs for larger systems (e.g., CN + squalene) using reactive pathways sampled in smaller systems (e.g., CN + isopentane). This will be interesting not only for experimental groups focusing on these systems, but also for the quantum machine learning community, given that NNs have yet to find extensive use in fitting reactive potential energy surfaces of large open shell systems.

## Acknowledgements

DRG acknowledges funding from the Royal Society as a University Research Fellow and also from EPSRC Program Grant EP/P021123/1. LAB acknowledges funding from EPSRC Program Grant EP/P021123/1. Funding for SA is from the EPSRC Centre for Doctoral training, Theory and Modelling in Chemical Sciences, under grant EP/L015722/1. We acknowledge MOC for helping to stabilize the software required to implement the quantum mechanical plugins for Narupa. SJB thanks EPSRC for grant EP/M022129/1, HECBioSim and the UoB School of Chemistry for funding. MR and ACV acknowledge financial support by ETH Zurich through grant ETH-20 15-1. The authors thank Mr. Stefan N. Heinen, University of Basel, for providing input files for the constrained MD. We further acknowledge the following programs: Open Babel[57], Pybel[58], Numpy[59], Matplotlib[60], VMD[61], Avogadro[62], OpenMP[40], F2PY.[63] We also acknowledge helpful conversations throughout with Prof. Matt Costen, Prof. Ken McKendrick and Dr. Stuart Greaves from Heriot-Watt University (Edinburgh) while preparing this draft.

# Supplementary Information

## S1. Hyper-parameter optimization

The Osprey code was modified to support Group K-Fold cross-validation. When using this technique, all structures from a given abstraction trajectory were part of either the training set OR test set, but they were not split across both. With standard K-Fold cross-validation one would risk overfitting, as the structures from a single abstraction trajectory are highly correlated (especially in the iMD-VR data set).

## S2. ACSFs re-parametrization

We also re-parametrized the original formulation of the ACSFs, to reduce the number of correlated hyper-parameters. As explained in the main text, the formulation of the ACSFs is:

$$g_2 = \sum_{j \neq i} e^{-\eta(R_{ij}-R_s)^2} f_c(R_{ij}) \tag{S1}$$

$$g_3 = 2^{1-\zeta} \sum_{j \neq i,k} (1 + \cos(\theta_{ijk} - \theta_s))^\zeta \times e^{-\eta\left(\frac{R_{ij}+R_{ik}}{2}-R_s\right)^2} f_c(R_{ij}) f_c(R_{ik}) \tag{S2}$$

We refer to $f(R)$ and $g(\theta)$ as the radial and angular 'basis functions':

$$f(R) = e^{-\eta(R-R_s)^2} \tag{S3}$$

$$g(\theta) = 2^{1-\zeta}(1 + \cos(\theta - \theta_s))^\zeta \tag{S4}$$

The hyper parameters $\eta$ and $\zeta$ control the width of the radial and angular basis functions respectively, while $R_s$ and $\theta_s$ control the location of their centers. If the values of $\eta$ and $\zeta$ are too large, the Gaussian functions will not overlap enough (Figure S13A), while if they are too large there will overlap excessively (Figure S13B). Usually, a grid of values of $R_s$ and $\theta_s$ is used to create the ACSFs. We use $N_r$ and $N_a$ to refer to the number of $R_s$ and $\theta_s$ values used in the grid. The values of $R_s$ range from $r_{min}$ to the cut-off radius $R_c$ and the values of $\theta_s$ range from 0 to $\pi$.

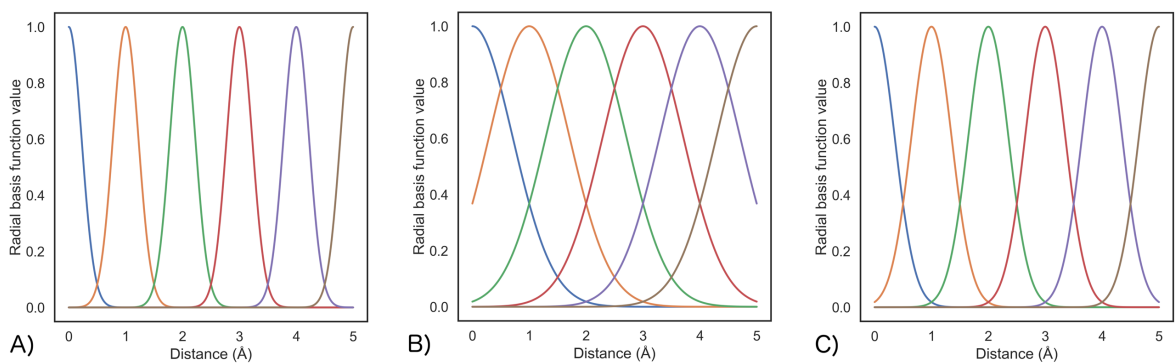

Figure S13 - Gaussians in the radial basis with different values of the $\eta$ parameter: (A) large value ($\eta = 10$), (B) small value ($\eta = 1$), (C) intermediate value ($\eta = 4$). The values of $R_s$ are [0, 1, 2, 3, 4, 5], so $r_{min} = 0$, $R_c = 5$ and $N_r = 6$.

To reduce the number of hyper-parameters, we use the same number of radial and angular basis functions (i.e. $N_r = N_a = N_{basis}$). Since choosing a good value for $\eta$ and $\zeta$ depends on the number of basis functions ($N_{basis}$), we re-write them as a function of $N_{basis}$ and a precision parameter $\tau$. We define $\tau$ such that the value where two neighbouring radial basis functions intersect is $1/\tau$ and the value where two neighbouring angular basis functions intersect is $2/\tau$. For $N_{basis}$ radial basis functions, with $R_s$ range from $r_{min}$ to $R_c$ (included), the distance between the centers of any two neighbouring basis function is

$$d = \frac{R_c - r_{min}}{N_{basis} - 1} \tag{S5}$$

It follows that neighbouring basis functions will intersect at a distance of $R_s + d/2$. This means that we can express $\eta$ as a function of $N_{basis}$, $R_c$, $r_{min}$ and $\tau$ by solving the equation

$$f\left(R_s + \frac{d}{2}\right) = \frac{1}{\tau} \tag{S6}$$

Resulting in

$$\eta = \frac{4 \cdot \log(\tau) \cdot (N_{basis} - 1)^2}{(R_c - r_{min})^2} \tag{S7}$$

Similarly, for $N_{basis}$ angular basis functions, with $\theta_s$ in the range 0 to $\pi$, the distance between any two neighbouring basis function centers is:

$$d = \frac{\pi}{N_{basis} - 1} \tag{S8}$$

It follows that a given basis function will intersect with a neighbouring one at $\theta_s + d/2$. This means that we can express $\zeta$ as a function of $N_{basis}$ and $\tau$ by solving the equation

$$g\left(\theta_s + \frac{d}{2}\right) = \frac{2}{\tau} \tag{S9}$$

Resulting in:

$$\zeta = -\frac{\log(\tau)}{2 \cdot \log\left(\cos\left(\frac{\pi}{4N_{basis} - 4}\right)\right)} \tag{S10}$$

*S3. Reproducing our results*

We have created a repository with instructions on how to reproduce our results. The repository can be found at https://github.com/SilviaAmAm/isopentane_paper_si

The repository contains:

- Input files to Molpro for running the constrained optimizations and single point energies and forces.
- Input files to CP2K for running the constrained MD.
- All the data sets that were used for the generation of the results in the paper.
- Scripts to fit the neural networks.
- Scripts for creating the plots.
- Scripts to use Osprey to optimize the hyper-parameters.
- A link to the video showing H-abstraction in Virtual Reality

# References

1.     Krizhevsky, A.; Sutskever, I.; Hinton, G. E., *ImageNet Classification with Deep Convolutional Neural Networks*. Curran Associates, Inc.: 2012.
2.     Silver, D.; Huang, A.; Maddison, C. J.; Guez, A.; Sifre, L.; van den Driessche, G.; Schrittwieser, J.; Antonoglou, I.; Panneershelvam, V.; Lanctot, M.; Dieleman, S.; Grewe, D.; Nham, J.; Kalchbrenner, N.; Sutskever, I.; Lillicrap, T.; Leach, M.; Kavukcuoglu, K.; Graepel, T.; Hassabis, D., Mastering the game of Go with deep neural networks and tree search. *Nature* **2016,** *529*, 484.
3.     Hinton, G.; Deng, L.; Yu, D.; Dahl, G. E.; Mohamed, A.; Jaitly, N.; Senior, A.; Vanhoucke, V.; Nguyen, P.; Sainath, T. N.; Kingsbury, B., Deep Neural Networks for Acoustic Modeling in Speech Recognition: The Shared Views of Four Research Groups. *IEEE Signal Processing Magazine* **2012,** *29* (6), 82-97.
4.     Auli, M.; Galley, M.; Quirk, C.; Zweig, G., Joint language and translation modeling with recurrent neural networks. **2013**.
5.     Curry, B.; Moutinho, L., Neural Networks in Marketing: Modelling Consumer Responses to Advertising Stimuli. *European Journal of Marketing* **1993,** *27* (7), 5-20.
6.     (a) Butler, K. T.; Davies, D. W.; Cartwright, H.; Isayev, O.; Walsh, A., Machine learning for molecular and materials science. *Nature* **2018,** *559* (7715), 547-555; (b) Rupp, M., Machine learning for quantum mechanics in a nutshell. *International Journal of Quantum Chemistry* **2015,** *115* (16), 1058-1073; (c) Bartók, A. P.; De, S.; Poelking, C.; Bernstein, N.; Kermode, J. R.; Csányi, G.; Ceriotti, M., Machine learning unifies the modeling of materials and molecules. *Science Advances* **2017,** *3* (12), e1701816; (d) Sanchez-Lengeling, B.; Aspuru-Guzik, A., Inverse molecular design using machine learning: Generative models for matter engineering. *Science* **2018,** *361* (6400), 360-365; (e) De, S.; Bartók, A. P.; Csányi, G.; Ceriotti, M., Comparing molecules and solids across structural and alchemical space. *Physical Chemistry Chemical Physics* **2016,** *18* (20), 13754-13769.
7.     Gupta, A.; Müller, A. T.; Huisman, B. J. H.; Fuchs, J. A.; Schneider, P.; Schneider, G., Generative Recurrent Networks for De Novo Drug Design. *Molecular Informatics* **2018,** *37* (1-2), 1700111.
8.     Olivares-Amaya, R.; Amador-Bedolla, C.; Hachmann, J.; Atahan-Evrenk, S.; Sánchez-Carrera, R. S.; Vogt, L.; Aspuru-Guzik, A., Accelerated computational discovery of high-performance materials for organic photovoltaics by means of cheminformatics. *Energy & Environmental Science* **2011,** *4* (12), 4849-4861.
9.     Segler, M. H. S.; Waller, M. P., Neural-Symbolic Machine Learning for Retrosynthesis and Reaction Prediction. *Chemistry – A European Journal* **2017,** *23* (25), 5966-5971.
10.    Mardt, A.; Pasquali, L.; Wu, H.; Noé, F., VAMPnets for deep learning of molecular kinetics. *Nature Communications* **2018,** *9* (1), 5.
11.    Braams, B. J.; Bowman, J. M., Permutationally invariant potential energy surfaces in high dimensionality. *International Reviews in Physical Chemistry* **2009,** *28* (4), 577-606.

12. Ischtwan, J.; Collins, M. A., Molecular potential energy surfaces by interpolation. *The Journal of Chemical Physics* **1994,** *100* (11), 8080-8088.
13. (a) Warshel, A.; M. Weiss, R., *An Empirical Valence Bond Approach for Comparing Reactions in Solutions and in Enzymes*. 1980; Vol. 102; (b) Glowacki, D. R.; Orr-Ewing, A. J.; Harvey, J. N., Non-equilibrium reaction and relaxation dynamics in a strongly interacting explicit solvent: F+ CD3CN treated with a parallel multi-state EVB model. *The Journal of chemical physics* **2015,** *143* (4), 044120.
14. Chmiela, S.; Tkatchenko, A.; Sauceda, H. E.; Poltavsky, I.; Schütt, K. T.; Müller, K.-R., Machine learning of accurate energy-conserving molecular force fields. *Science Advances* **2017,** *3* (5).
15. Ho, T.-S.; Rabitz, H., Reproducing kernel Hilbert space interpolation methods as a paradigm of high dimensional model representations: Application to multidimensional potential energy surface construction. *The Journal of Chemical Physics* **2003,** *119* (13), 6433-6442.
16. Behler, J., Neural network potential-energy surfaces for atomistic simulations. In *Chemical Modelling: Applications and Theory Volume 7*, The Royal Society of Chemistry: 2010; Vol. 7, pp 1-41.
17. Pukrittayakamee, A.; Malshe, M.; Hagan, M.; Raff, L. M.; Narulkar, R.; Bukkapatnum, S.; Komanduri, R., Simultaneous fitting of a potential-energy surface and its corresponding force fields using feedforward neural networks. *The Journal of Chemical Physics* **2009,** *130* (13), 134101.
18. Guàrdia, E.; Rey, R.; Padró, J. A., Potential of mean force by constrained molecular dynamics: A sodium chloride ion-pair in water. *Chemical Physics* **1991,** *155* (2), 187-195.
19. Bernardi, R. C.; Melo, M. C. R.; Schulten, K., Enhanced sampling techniques in molecular dynamics simulations of biological systems. *Biochimica et biophysica acta* **2015,** *1850* (5), 872-877.
20. Collins, M. A., Molecular potential-energy surfaces for chemical reaction dynamics. *Theoretical Chemistry Accounts* **2002,** *108* (6), 313-324.
21. Csányi, G.; Albaret, T.; Payne, M. C.; De Vita, A., ``Learn on the Fly'': A Hybrid Classical and Quantum-Mechanical Molecular Dynamics Simulation. *Physical Review Letters* **2004,** *93* (17), 175503.
22. (a) Greaves, S. J.; Rose, R. A.; Oliver, T. A. A.; Glowacki, D. R.; Ashfold, M. N. R.; Harvey, J. N.; Clark, I. P.; Greetham, G. M.; Parker, A. W.; Towrie, M.; Orr-Ewing, A. J., Vibrationally Quantum-State–Specific Reaction Dynamics of H Atom Abstraction by CN Radical in Solution. *Science* **2011,** *331* (6023), 1423; (b) Glowacki, D. R.; Orr-Ewing, A. J.; Harvey, J. N., Product energy deposition of CN + alkane H abstraction reactions in gas and solution phases. *The Journal of Chemical Physics* **2011,** *134* (21), 214508; (c) Dunning, G.; Glowacki, D.; Preston, T.; Greaves, S.; Greetham, G.; Clark, I.; Towrie, M.; Harvey, J.; Orr-Ewing, A., Vibrational relaxation and microsolvation of DF after F-atom reactions in polar solvents. *Science* **2015,** *347* (6221), 530-533; (d) Carpenter, B. K.; Harvey, J. N.; Glowacki, D. R., Prediction of enhanced solvent-induced enantioselectivity for a ring opening with a bifurcating reaction path. *Physical Chemistry Chemical Physics* **2015,** *17* (13), 8372-8381.
23. O'Connor, M.; Deeks, H. M.; Dawn, E.; Metatla, O.; Roudaut, A.; Sutton, M.; Thomas, L. M.; Glowacki, B. R.; Sage, R.; Tew, P.; Wonnacott, M.; Bates, P.; Mulholland, A. J.; Glowacki, D. R., Sampling molecular conformations and dynamics in a multiuser virtual reality framework. *Science Advances* **2018,** *4* (6).
24. (a) Chen, J.; Xu, X.; Zhang, D. H., Communication: An accurate global potential energy surface for the OH + CO → H + CO2 reaction using neural networks. *The Journal of Chemical Physics* **2013,** *138* (22), 221104; (b) Li, J.; Jiang, B.; Guo, H., Permutation invariant polynomial neural network approach to fitting potential energy surfaces. II. Four-atom systems. *The Journal of Chemical Physics* **2013,** *139* (20), 204103.
25. Abadi, M.; Barham, P.; Chen, J.; Chen, Z.; Davis, A.; Dean, J.; Devin, M.; Ghemawat, S.; Irving, G.; Isard, M.; Kudlur, M.; Levenberg, J.; Monga, R.; Moore, S.; Murray, D. G.; Steiner, B.; Tucker, P.; Vasudevan, V.; Warden, P.; Wicke, M.; Yu, Y.; Zheng, X., TensorFlow: A System for Large-Scale Machine Learning. 12th USENIX Symposium on Operating Systems Design and Implementation, 2016.
26. Paszke, A.; Gross, S.; Chintala, S.; Chanan, G.; Yang, E.; DeVito, Z.; Lin, Z.; Desmaison, A.; Antiga, L.; Lerer, A., Automatic differentiation in PyTorch. NIPS-W, 2017.
27. (a) Heck, R.; Vuculescu, O.; Sørensen, J. J.; Zoller, J.; Andreasen, M. G.; Bason, M. G.; Ejlertsen, P.; Elíasson, O.; Haikka, P.; Laustsen, J. S.; Nielsen, L. L.; Mao, A.; Müller, R.; Napolitano, M.; Pedersen, M. K.; Thorsen, A. R.; Bergenholtz, C.; Calarco, T.; Montangero, S.; Sherson, J. F., Remote optimization of an ultracold atoms experiment by experts and citizen scientists. *Proceedings of the National Academy of Sciences* **2018,** *115* (48), E11231; (b) Cooper, S.; Khatib, F.; Treuille, A.; Barbero, J.; Lee, J.; Beenen, M.; Leaver-Fay, A.; Baker, D.; Popović, Z.; players, F., Predicting protein structures with a multiplayer online game. *Nature* **2010,** *466*, 756; (c) Khatib, F.; Cooper, S.; Tyka, M. D.; Xu, K.; Makedon, I.; Popović, Z.; Baker, D.; Players, F., Algorithm discovery by protein folding game players. *Proceedings of the National Academy of Sciences* **2011**; (d) Cooper, S.; Treuille, A.; Barbero, J.; Leaver-Fay, A.; Tuite, K.; Khatib, F.; Snyder, A. C.; Beenen, M.; Salesin, D.; Baker, D. In *The challenge of designing scientific discovery games*, Proceedings of the Fifth international Conference on the Foundations of Digital Games, ACM: 2010; pp 40-47; (e) Horowitz, S.; Koepnick, B.; Martin, R.; Tymieniecki, A.; Winburn, A. A.; Cooper, S.; Flatten, J.; Rogawski, D. S.; Koropatkin, N. M.; Hailu, T. T., Determining crystal structures through crowdsourcing and coursework. *Nature communications* **2016,** *7*, 12549; (f) Khoury, G. A.; Liwo, A.; Khatib, F.; Zhou, H.; Chopra, G.; Bacardit, J.; Bortot, L. O.; Faccioli, R. A.; Deng, X.; He, Y.; Krupa, P.; Li, J.; Mozolewska, M. A.; Sieradzan, A. K.; Smadbeck, J.; Wirecki, T.; Cooper, S.; Flatten, J.; Xu, K.; Baker, D.; Cheng, J.; Delbem, A. C. B.; Floudas, C. A.; Keasar, C.; Levitt, M.; Popović, Z.; Scheraga, H. A.; Skolnick, J.; Crivelli, S. N.; Players, F., WeFold: A coopetition for protein structure prediction. *Proteins: Structure, Function, and Bioinformatics* **2014,** *82* (9), 1850-1868.
28. Behler, J., Constructing high-dimensional neural network potentials: A tutorial review. *International Journal of Quantum Chemistry* **2015,** *115* (16), 1032-1050.
29. Schütt, K. T.; Sauceda, H. E.; Kindermans, P.-J.; Tkatchenko, A.; Müller, K.-R., SchNet – A deep learning architecture for molecules and materials. *The Journal of Chemical Physics* **2018,** *148* (24), 241722.
30. Yao, K.; Herr, J. E.; Toth, David W.; McKintyre, R.; Parkhill, J., The TensorMol-0.1 model chemistry: a neural network augmented with long-range physics. *Chemical Science* **2018,** *9* (8), 2261-2269.
31. Smith, J. S.; Isayev, O.; Roitberg, A. E., ANI-1: an extensible neural network potential with DFT accuracy at force field computational cost. *Chemical Science* **2017,** *8* (4), 3192-3203.
32. Khorshidi, A.; Peterson, A. A., Amp: A modular approach to machine learning in atomistic simulations. *Computer Physics Communications* **2016,** *207*, 310-324.
33. Behler, J., Atom-centered symmetry functions for constructing high-dimensional neural network potentials. *The Journal of Chemical Physics* **2011,** *134* (7), 074106.
34. (a) Pedregosa, F.; Varoquaux, G.; Gramfort, A.; Michel, V.; Thirion, B.; Grisel, O.; Blondel, M.; Prettenhofer, P.; Weiss, R.; Dubourg, V., Scikit-learn: Machine learning in Python. *Journal of machine learning research* **2011,** *12* (Oct), 2825-2830; (b) Buitinck, L.; Louppe, G.; Blondel, M.; Pedregosa, F.; Mueller, A.; Grisel, O.; Niculae, V.; Prettenhofer, P.; Gramfort, A.; Grobler, J., API design for machine learning software: experiences from the scikit-learn project. *arXiv preprint arXiv:1309.0238* **2013**.
35. McGibbon, R. T.; Hernández, C. X.; Harrigan, M. P.; Kearnes, S.; Sultan, M. M.; Jastrzebski, S.; Husic, B. E.; Pande, V. S., Osprey: Hyperparameter Optimization for Machine Learning. The Journal of Open Source Software: 2016.
36. Christensen, A. S.; Bratholm, L. A.; Amabilino, S.; Kromann, J. C.; Faber, F. A.; Huang, B.; Glowacki, D. R.; Tkatchenko, A.; Muller, K. R.; von Lilienfeld, O. A. QML: A Python Toolkit for Quantum Machine Learning. http://www.qmlcode.org.
37. (a) Bartók, A. P.; Kondor, R.; Csányi, G., On representing chemical environments. *Physical Review B* **2013,** *87* (18), 184115; (b) Behler, J.; Parrinello, M., Generalized Neural-Network Representation of High-Dimensional Potential-Energy Surfaces. *Physical Review Letters* **2007,** *98* (14), 146401.
38. Rupp, M.; Tkatchenko, A.; Müller, K.-R.; von Lilienfeld, O. A., Fast and Accurate Modeling of Molecular Atomization Energies with Machine Learning. *Physical Review Letters* **2012,** *108* (5), 058301.
39. Huang, B.; Anatole von Lilienfeld, O., The "DNA" of chemistry: Scalable quantum machine learning with "amons". *eprint arXiv:1707.04146* **2017**, arXiv:1707.04146.
40. Dagum, L.; Menon, R., OpenMP: an industry standard API for shared-memory programming. *IEEE computational science and engineering* **1998,** *5* (1), 46-55.
41. (a) Moćkus, J.; Tiesis, V.; Źilinskas, A., The Application of Bayesian Methods for Seeking the Extremum. Vol. 2. Elsevier: 1978; (b) Jones, D. R.; Schonlau, M.; Welch, W. J., Efficient Global Optimization of Expensive Black-Box Functions. *Journal of Global Optimization* **1998,** *13* (4), 455-492.
42. Auer, P.; Cesa-Bianchi, N.; Fischer, P., Finite-time analysis of the multiarmed bandit problem. *Machine learning* **2002,** *47* (2-3), 235-256.
43. (a) Akaike, H., A new look at the statistical model identification. *IEEE Transactions on Automatic Control* **1974,** *19* (6), 716-723; (b) Burnham, K. P.; Anderson, D. R., *Model selection and multimodel inference: a practical information-theoretic approach*. Springer Science & Business Media: 2003; (c) Cavanaugh, J. E., Unifying the derivations for the Akaike and corrected Akaike information criteria. *Statistics & Probability Letters* **1997,** *33* (2), 201-208.
44. Aradi, B.; Hourahine, B.; Frauenheim, T., DFTB+, a Sparse Matrix-Based Implementation of the DFTB Method. *The Journal of Physical Chemistry A* **2007,** *111* (26), 5678-5684.
45. (a) Haag, M. P.; Reiher, M., Real-time quantum chemistry. *International Journal of Quantum Chemistry* **2012,** *113* (1), 8-20; (b) Vaucher, A. C.; Haag, M. P.; Reiher, M., Real-time feedback from iterative electronic structure calculations. *Journal of Computational Chemistry* **2016,** *37* (9), 805-812; (c) Haag, M. P.; Reiher, M., Studying chemical reactivity in a virtual environment. *Faraday Discussions* **2014,** *169* (0), 89-118.

46. Husch, T.; Vaucher, A. C.; Reiher, M., Semiempirical molecular orbital models based on the neglect of diatomic differential overlap approximation. *International Journal of Quantum Chemistry* **2018,** *118* (24), e25799.
47. Stewart, J. J. P., Optimization of parameters for semiempirical methods V: Modification of NDDO approximations and application to 70 elements. *Journal of Molecular Modeling* **2007,** *13* (12), 1173-1213.
48. (a) Hutter, J.; Iannuzzi, M.; Schiffmann, F.; VandeVondele, J., cp2k: atomistic simulations of condensed matter systems. *Wiley Interdisciplinary Reviews: Computational Molecular Science* **2014,** *4* (1), 15-25; (b) Borštnik, U.; VandeVondele, J.; Weber, V.; Hutter, J., Sparse matrix multiplication: The distributed block-compressed sparse row library. *Parallel Computing* **2014,** *40* (5), 47-58; (c) Frigo, M.; Johnson, S. G., The Design and Implementation of FFTW3. *Proceedings of the IEEE* **2005,** *93* (2), 216-231; (d) Kolafa, J., Time-reversible always stable predictor–corrector method for molecular dynamics of polarizable molecules. *Journal of Computational Chemistry* **2004,** *25* (3), 335-342.
49. Bussi, G.; Donadio, D.; Parrinello, M., Canonical sampling through velocity rescaling. *The Journal of Chemical Physics* **2007,** *126* (1), 014101.
50. (a) Werner, H.-J.; Knowles, P. J.; Knizia, G.; Manby, F. R.; Schütz, M., Molpro: a general-purpose quantum chemistry program package. *Wiley Interdisciplinary Reviews: Computational Molecular Science* **2012,** *2* (2), 242-253; (b) Werner, H.-J.; Knowles, P. J.; Knizia, G.; Manby, F. R.; Schütz, M.; Celani, P.; Györffy, W.; Kats, D.; Korona, T.; Lindh, R.; Mitrushenkov, A.; Rauhut, G.; Shamasundar, K. R.; Adler, T. B.; Amos, R. D.; Bennie, S. J.; Bernhardsson, A.; Berning, A.; Cooper, D. L.; Deegan, M. J. O.; Dobbyn, A. J.; Eckert, F.; Goll, E.; Hampel, C.; Hesselmann, A.; Hetzer, G.; Hrenar, T.; Jansen, G.; Köppl, C.; Lee, S. J. R.; Liu, Y.; Lloyd, A. W.; Ma, Q.; Mata, R. A.; May, A. J.; McNicholas, S. J.; Meyer, W.; Miller III, T. F.; Mura, M. E.; Nicklaß, A.; O'Neill, D. P.; Palmieri, P.; Peng; Pflüger, K.; Pitzer, R.; Reiher, M.; Shiozaki, T.; Stoll, H.; Stone, A. J.; Tarroni, R.; Thorsteinsson, T.; Wang, M.; Welbor, M., MOLPRO, version 2018.2, a package of ab initio programs. *There is no corresponding record for this reference*; (c) Lindh, R., The reduced multiplication scheme of the Rys-Gauss quadrature for 1st order integral derivatives. *Theoretica chimica acta* **1993,** *85* (6), 423-440; (d) Lindh, R.; Ryu, U.; Liu, B., The reduced multiplication scheme of the Rys quadrature and new recurrence relations for auxiliary function based two‐electron integral evaluation. *The Journal of chemical physics* **1991,** *95* (8), 5889-5897; (e) Eckert, F.; Pulay, P.; Werner, H. J., Ab initio geometry optimization for large molecules. *Journal of computational chemistry* **1997,** *18* (12), 1473-1483.
51. Polly, R.; Werner, H.-J.; Manby, F. R.; Knowles, P. J., Fast Hartree–Fock theory using local density fitting approximations. *Molecular Physics* **2004,** *102* (21-22), 2311-2321.
52. Perdew, J. P.; Burke, K.; Ernzerhof, M., Generalized Gradient Approximation Made Simple [Phys. Rev. Lett. 77, 3865 (1996)]. *Physical Review Letters* **1997,** *78* (7), 1396-1396.
53. (a) Weigend, F.; Ahlrichs, R., Balanced basis sets of split valence, triple zeta valence and quadruple zeta valence quality for H to Rn: Design and assessment of accuracy. *Physical Chemistry Chemical Physics* **2005,** *7* (18), 3297-3305; (b) Weigend, F., Accurate Coulomb-fitting basis sets for H to Rn. *Physical Chemistry Chemical Physics* **2006,** *8* (9), 1057-1065.
54. Hess, W. P.; Durant, J. L.; Tully, F. P., Kinetic study of the reactions of cyanogen radical with ethane and propane. *The Journal of Physical Chemistry* **1989,** *93* (17), 6402-6407.
55. Adamo, C.; Barone, V., Toward reliable density functional methods without adjustable parameters: The PBE0 model. *The Journal of chemical physics* **1999,** *110* (13), 6158-6170.
56. (a) Tesa-Serrate, M. A.; King, K. L.; Paterson, G.; Costen, M. L.; McKendrick, K. G., Site and bond-specific dynamics of reactions at the gas–liquid interface. *Physical Chemistry Chemical Physics* **2014,** *16* (1), 173-183; (b) Köhler, S. P. K.; Reed, S. K.; Westacott, R. E.; McKendrick, K. G., Molecular Dynamics Study to Identify the Reactive Sites of a Liquid Squalane Surface. *The Journal of Physical Chemistry B* **2006,** *110* (24), 11717-11724.
57. O'Boyle, N. M.; Banck, M.; James, C. A.; Morley, C.; Vandermeersch, T.; Hutchison, G. R., Open Babel: An open chemical toolbox. *Journal of cheminformatics* **2011,** *3* (1), 33.
58. O'Boyle, N. M.; Morley, C.; Hutchison, G. R., Pybel: a Python wrapper for the OpenBabel cheminformatics toolkit. *Chemistry Central Journal* **2008,** *2* (1), 5.
59. Oliphant, T. E., *A guide to NumPy*. Trelgol Publishing USA: 2006; Vol. 1.
60. Hunter, J. D., Matplotlib: A 2D Graphics Environment. *Computing in Science & Engineering* **2007,** *9* (3), 90-95.
61. Humphrey, W.; Dalke, A.; Schulten, K., VMD: visual molecular dynamics. *Journal of molecular graphics* **1996,** *14* (1), 33-38.
62. (a) Avogadro: an open-source molecular builder and visualization tool. Version 1.2.0. http://avogadro.cc/; (b) Hanwell, M. D.; Curtis, D. E.; Lonie, D. C.; Vandermeersch, T.; Zurek, E.;

Hutchison, G. R., Avogadro: an advanced semantic chemical editor, visualization, and analysis platform. *Journal of Cheminformatics* **2012,** *4* (1), 17.
63. Peterson, P., F2PY: a tool for connecting Fortran and Python programs. *International Journal of Computational Science and Engineering* **2009,** *4* (4), 296-305.